\documentclass[amsmath,amsfonts,amssymb,aps,prx,twocolumn,superscriptaddress,citeautoscript,longbibliography]{revtex4-2}
\usepackage{graphicx}
\usepackage{xcolor}
\usepackage[colorlinks=true,citecolor=orange,linkcolor=blue]{hyperref}

\begin{document}
\title{Probing critical behavior of long-range transverse-field Ising model through quantum Kibble-Zurek mechanism}
\author{B.-W. Li}
\thanks{These authors contribute equally to this work}%
\affiliation{Center for Quantum Information, Institute for Interdisciplinary Information Sciences, Tsinghua University, Beijing 100084, PR China}
\affiliation{Hefei National Laboratory, Hefei 230088, PR China}

\author{Y.-K. Wu}
\thanks{These authors contribute equally to this work}%
\affiliation{Center for Quantum Information, Institute for Interdisciplinary Information Sciences, Tsinghua University, Beijing 100084, PR China}
\affiliation{Hefei National Laboratory, Hefei 230088, PR China}

\author{Q.-X. Mei}
\thanks{These authors contribute equally to this work}%
\affiliation{Center for Quantum Information, Institute for Interdisciplinary Information Sciences, Tsinghua University, Beijing 100084, PR China}
\affiliation{HYQ Co., Ltd., Beijing 100176, PR China}

\author{R. Yao}
\affiliation{Center for Quantum Information, Institute for Interdisciplinary Information Sciences, Tsinghua University, Beijing 100084, PR China}
\affiliation{Hefei National Laboratory, Hefei 230088, PR China}

\author{W.-Q. Lian}
\affiliation{Center for Quantum Information, Institute for Interdisciplinary Information Sciences, Tsinghua University, Beijing 100084, PR China}
\affiliation{HYQ Co., Ltd., Beijing 100176, PR China}

\author{M.-L. Cai}
\affiliation{HYQ Co., Ltd., Beijing 100176, PR China}

\author{Y. Wang}
\affiliation{Center for Quantum Information, Institute for Interdisciplinary Information Sciences, Tsinghua University, Beijing 100084, PR China}
\affiliation{Hefei National Laboratory, Hefei 230088, PR China}

\author{B.-X. Qi}
\affiliation{Center for Quantum Information, Institute for Interdisciplinary Information Sciences, Tsinghua University, Beijing 100084, PR China}
\affiliation{Hefei National Laboratory, Hefei 230088, PR China}

\author{L. Yao}
\affiliation{HYQ Co., Ltd., Beijing 100176, PR China}

\author{L. He}
\affiliation{Center for Quantum Information, Institute for Interdisciplinary Information Sciences, Tsinghua University, Beijing 100084, PR China}
\affiliation{Hefei National Laboratory, Hefei 230088, PR China}

\author{Z.-C. Zhou}
\affiliation{Center for Quantum Information, Institute for Interdisciplinary Information Sciences, Tsinghua University, Beijing 100084, PR China}
\affiliation{Hefei National Laboratory, Hefei 230088, PR China}

\author{L.-M. Duan}
\email{lmduan@tsinghua.edu.cn}
\affiliation{Center for Quantum Information, Institute for Interdisciplinary Information Sciences, Tsinghua University, Beijing 100084, PR China}
\affiliation{Hefei National Laboratory, Hefei 230088, PR China}

\begin{abstract}
The trapped ion quantum simulator has demonstrated qualitative properties of different physical models for up to tens of ions. In particular, a linear ion chain naturally hosts long-range Ising interactions under the laser driving, which has been used for various phenomena such as quantum phase transition, localization, thermalization and information propagation. For near-term practical usage, a central task is to find more quantitative applications of the noisy quantum simulators that are robust to small errors in the parameters. Here we report the quantum simulation of a long-range transverse-field Ising model using up to 61 ions and probe the critical behavior of its quantum phase transition through the Kibble-Zurek mechanism. By calibrating and verifying the coupling coefficients, we realize the same model for increasing ion numbers, so as to extract a critical exponent free of the finite size effect. For ferromagnetic interaction, our experimental result agrees with the previous numerical prediction. As for the anti-ferromagnetic case, signals are too weak to fit a critical exponent due to the frustration in the interaction, but still consistent with the theory.
\end{abstract}

\maketitle

\section{Introduction}
The transverse-field Ising model \cite{PFEUTY197079} is an iconic model in quantum many-body physics, and is widely used to illustrate various equilibrium and dynamical properties such as quantum phase transition \cite{sachdev_2011}, spin glass \cite{RevModPhys.58.801} and many-body localization \cite{RevModPhys.91.021001}. In the one-dimensional (1D) case with nearest-neighbor interaction, this model can be analytically solved by mapping to free fermions through Jordan-Wigner transformation \cite{PFEUTY197079}. However, when long range interaction is considered \cite{PhysRevB.64.184106,PhysRevLett.106.130601,PhysRevLett.109.267203,PhysRevLett.111.207202,PhysRevX.3.031015,Vodola_2015,PhysRevB.94.075156,PhysRevB.96.134427}, the problem becomes more complicated and generally relies on numerical approaches like quantum Monte Carlo \cite{PhysRevB.43.5950} or tensor-network-based methods \cite{schollwock2011densitymatrix}.

As one of the leading platforms for quantum information processing, the trapped ion system naturally supports the long-range Ising-type interaction under laser driving \cite{PhysRevLett.92.207901,PhysRevLett.103.120502,monroe2021Programmable}, thus is convenient for the quantum simulation of this model. Previous experiments have demonstrated various aspects of the long-range transverse-field Ising model such as equilibrium and dynamical phase transitions \cite{islam2011Onset,zhang2017Observation}, localization \cite{smith2016many,doi:10.1126/science.aau4963}, thermalization \cite{doi:10.1126/sciadv.1700672} and information propagation \cite{richerme2014non,jurcevic2014quasiparticle}. However, since quantum error correction is still not available in this noisy intermediate-scale quantum (NISQ) era \cite{Preskill2018quantumcomputingin}, current multi-ion experiments are subjected to considerable experimental noise and errors. Therefore these previous demonstrations mainly focus on the qualitative features, while quantitative analysis in the trapped-ion-based quantum simulators is still largely lacking. On the other hand, some universal properties, such as the critical exponents of a class of quantum phase transition \cite{sachdev_2011}, are more robust to experimental noise and shall survive in the NISQ devices. These properties thus make ideal candidates for the quantitative applications of the ion trap quantum simulator.

The Kibble-Zurek mechanism (KZM) \cite{kibble1976topology,kibble1980some,zurek1985cosmological,zurek1996cosmological} provides a possible scheme to probe the critical behavior of the quantum phase transition through a slow quench across the phase transition point \cite{zurek2005dynamics,PhysRevLett.95.035701,dziarmaga2005dynamics,polkovnikov2005universal}. Roughly speaking, as we approach the critical point from one phase, the system stays in the ground state adiabatically until we are sufficiently close to the phase transition point where the energy gap closes in the thermodynamic limit and thus the adiabatic condition breaks down. During this intermediate region, the evolution of the system approximately freezes out so that excitations (defects) are created when the system leaves the critical region and the evolution again becomes adiabatic. The defect density $\rho$ is determined by how close we are to the critical point when leaving the adiabatic region, thus a power-law scaling versus the quench time $\rho \propto T^{-\mu}$. This gives us a critical exponent $\mu$ of the quantum phase transition and is insensitive to the experimental imperfections.

A great deal of efforts have been made to test KZM in classical \cite{PhysRevLett.89.080603,ulm2013observation,pyka2013topological,doi:10.1126/science.1258676} and quantum \cite{doi:10.1126/science.aaf9657,PhysRevLett.116.155301,keesling2019quantum} phase transitions. For the transverse-field Ising model with long-range interaction, quantum KZM has been proposed to study its critical behavior \cite{jaschke2017critical,puebla2019quantum} and pioneering works in Rydberg atoms have been demonstrated \cite{keesling2019quantum}. However, for Rydberg atoms, the Ising-type coupling quickly decays as $r^{-6}$ such that the transition belongs to the same universality class as the nearest-neighbor interaction \cite{PhysRevLett.109.267203,PhysRevB.94.075156,puebla2019quantum}. On the other hand, in ion trap it is more convenient to achieve $r^{-\alpha}$ interaction with $\alpha\in[0.5,1.8]$ \cite{zhang2017Observation} (and in principle for any $\alpha\in[0,3]$\cite{PhysRevLett.92.207901,islam2011Onset}) where interesting physics such as an $\alpha$-dependent critical exponent is predicted \cite{jaschke2017critical,puebla2019quantum}.

Here we report the quantum simulation of the 1D long-range transverse-field Ising model using up to 61 ions. After calibrating and verifying the Ising coupling coefficients, we probe the critical behavior of the ferromagnetic (FM) model through the KZM and study its scaling with the ion number. In particular, we obtain a shared critical exponent for sufficiently large ion numbers, hence ruling out the boundary effects for the finite system (a more detailed finite-size scaling can be found in Appendix~\ref{app:theoretical}). As for the anti-ferromagnetic (AFM) case, we still get consistent results with the theoretical prediction, but it is more difficult to approach the critical region due to the frustration and thus much smaller energy gap of the Hamiltonian.

\begin{figure*}[!tbp]
	\centering
	\includegraphics[width=\textwidth]{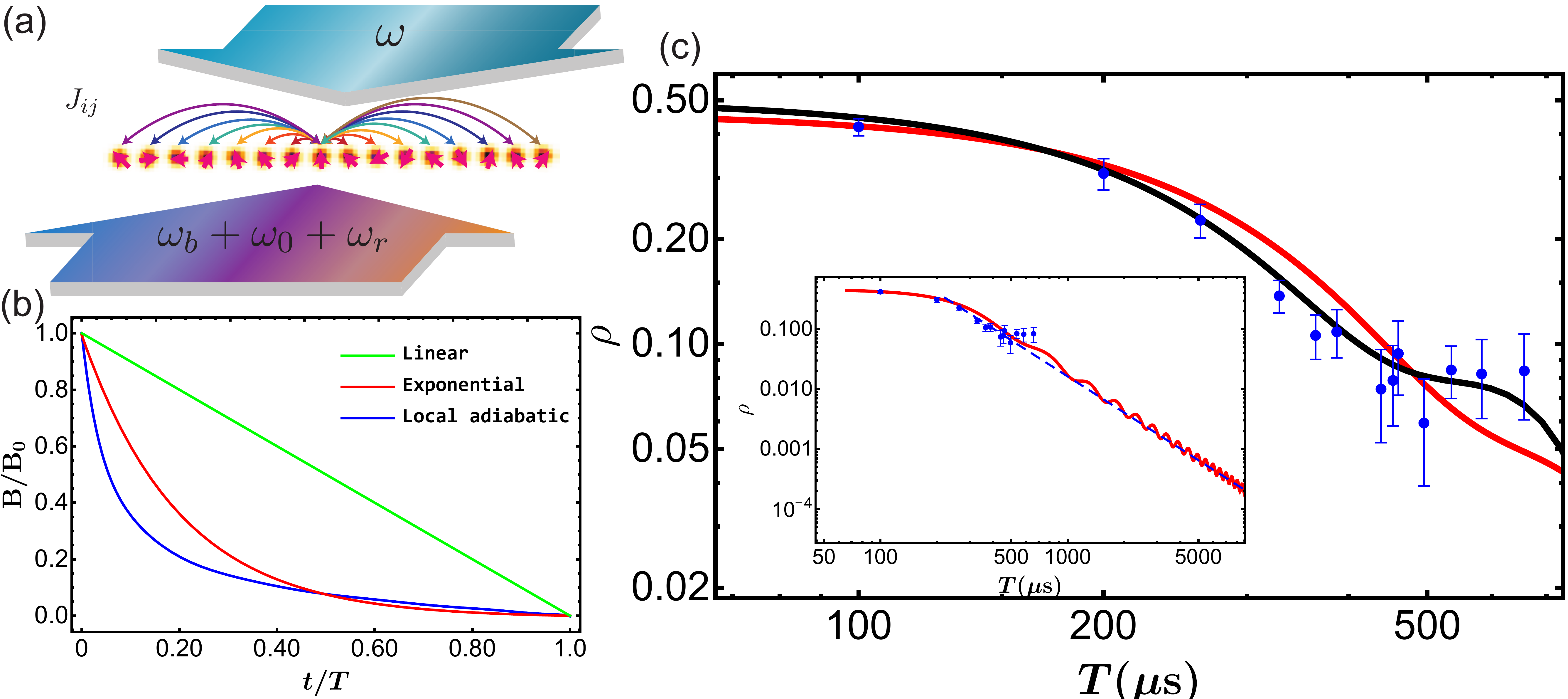}
	\caption {Experimental scheme. (a) The Ising model Hamiltonian is achieved on a chain of up to 61 ions using two counter-propagating $355\,$nm global Raman laser beams. (b) There are different paths to quench the transverse field. The original Kibble-Zurek mechanism (KZM) assumes a linear quench (green). In this experiment, restricted by the coherence time of the ions under strong driving laser, a local adiabatic path (blue) can be used to optimize the adiabatic condition over the whole ramping dynamics, thus allowing us to enter the adiabatic regime as shown in (c). Later when probing the KZM for multiple ions, we take an exponential path (red) for both theoretical and experimental convenience.
(c) To illustrate different regions of the quench speed, we consider two ions with ferromagnetic coupling $J_0 = - 2\pi\times 450\,$Hz. Starting from a fully polarized state $|\uparrow_y\uparrow_y\rangle$ in the $y$ direction which is close to the ground state under $B_0 = 6 |J_0|$, we ramp down $B(t)$ following the local adiabatic path. For sufficiently low quench time, the defect density $\rho$ is close to $1/2$, namely a random and independent distribution $\sigma_x^i=\pm 1$ for the two ions. As the quench time increases, the defect density decays as more and more population remains in the ferromagnetic ground states. The blue dots with error bars representing one standard deviation are the experimental data, which agree well with the theoretical results (solid black curve). It deviates slightly from the solid red curve, which is the theoretical result when starting from the true ground state under $B_0$ rather than the fully polarized state. As shown in the inset, for the two-ion case with a large energy gap, we are just able to enter the adiabatic regime where the defect density follows $\rho\propto T^{-2}$ (dashed line). For larger ion numbers and thus smaller energy gaps, we will not be able to dive deep into the adiabatic regime, but the critical behavior can still be probed by the KZM.
    \label{fig1}}
\end{figure*}

\section{Experimental scheme}
Our experimental system consists of a linear chain of trapped $^{171}\mathrm{Yb}^+$ ions as shown in Fig.~\ref{fig1}(a). The spin states are encoded in the clock states $|F=0,m_F=0\rangle\equiv |\downarrow_z \rangle$ and $|F=1,m_F=0\rangle\equiv |\uparrow_z \rangle$ of the $\mathrm{S}_{1/2}$ manifold separated by $\omega_{\mathrm{HF}} \approx 2\pi\times 12.64\,$GHz. The spin-spin coupling can be realized by bichromatic Raman laser beams with beatnotes $\omega_{\mathrm{HF}}\pm\delta$ \cite{PhysRevLett.92.207901,PhysRevLett.103.120502,monroe2021Programmable}, where we set the frequency detuning $\delta$ sufficiently away from all the motional sidebands to suppress the phonon excitation. In addition, we drive the carrier transition with a phase shift of $\pi/2$ to generate a transverse field. Together, we get a long-range transverse-field Ising model Hamiltonian
\begin{equation}
H = \sum_{i<j} J_{ij}\sigma_x^i \sigma_x^j +\sum_i B_i\sigma_y^i,\label{eq1}
\end{equation}
where $\sigma_{x(y)}^{i}$ are the Pauli operators on the ion $i$ and the summation $\sum_{i<j}$ runs over all ion pairs. $J_{ij}=\Omega_i \Omega_j \sum_k \eta_k^2 b_{ik}b_{jk}\omega_k/(\delta^2-\omega_k^2)$ is the effective spin-spin coupling depending on the laser detuning $\delta$, the Rabi frequency on each ion $\Omega_i$ and the phonon modes with frequency $\omega_k$, Lamb-Dicke parameter $\eta_k$ and mode vectors $b_{ik}$. Here the Rabi frequencies $\Omega_i$ and $B_i$ can be site-dependent because of the nonuniform laser intensity over the long ion chain. Typically the Ising coupling can be approximated by $J_{ij}\approx J_0/|i-j|^\alpha$ with $\alpha\in [0.5,1.8]$ \cite{zhang2017Observation} (which is however imperfect as we further discuss in Appendix~\ref{app:theoretical}). In this work, we achieve $\alpha\approx 1$ and an AFM coupling $J_0>0$ for various ion numbers up to 61. To simulate the FM interaction, we initialize the system in the highest eigenstate, which is effectively the ground state of the Hamiltonian $-H$. In the following, we will mainly focus on the FM model because it gives stronger experimental signals, and will postpone the discussion about the AFM model to the end of the paper.

It is well-known that the transverse-field Ising model possesses two phases with a quantum phase transition in between. At large transverse field $B\gg |J_0|$, the system is in the paramagnetic phase where the spins align with the field in the ground state; when $B\to 0$, we have double-degenerate FM ground states with all the spins aligned in either $+x$ or $-x$ directions. As we tune the transverse field continuously, a quantum phase transition occurs at a second order phase transition point. Near this point, the energy gap shrinks and the correlation length diverges in the thermodynamic limit, so that one can expect universal critical behavior insensitive to the microscopic structure of the model such as small fluctuation in the parameters.

The quantum KZM can be used to measure this critical behavior. Near the critical point, the temporal correlation length diverges as $c_0|g-g_c|^{-z\nu}$ where $c_0$ is a constant, $g$ an external parameter, e.g. the transverse field we consider here, $g_c$ the critical point, and $z$ and $\nu$ two critical exponents. This deviation $|g-g_c|$ from the critical point is achieved in a timescale $c_1|g-g_c|T$ if we scan the external parameter linearly with time for a total evolution time $T$, where $c_1$ is another constant. Combining these two scalings together, we obtain the boundary between the adiabatic and the ``frozen-out'' regimes in the KZM as $|g-g_c|\propto T^{-1/(1+z\nu)}$. This further gives us the spatial correlation length after the slow quench as $|g-g_c|^{-\nu}\propto T^{\nu/(1+z\nu)}\equiv T^{\mu}$, and hence a defect density $\rho\propto T^{-\mu}$ where we define \cite{jaschke2017critical}
\begin{equation}
\rho = \frac{1}{2(N-1)}\sum_i^{N-1}(1-\langle\sigma_x^i\sigma_x^{i+1}\rangle).
\end{equation}

In the above derivation, we assume a linear quench of the transverse field. However, since the KZM is only affected by the close neighborhood of the critical point, other forms of $B(t)$ also work if they can be approximated as linear in the ``frozen-out'' regime. In particular, current NISQ experiments are restricted by the finite coherence time of the system, so that we want to shorten the total evolution time by performing fast quench away from the critical point and only slowing down when the energy gap $\Delta$ shrinks. We can therefore use exponential or local adiabatic paths to quench the transverse field as shown in Fig.~\ref{fig1}(b). In Fig.~\ref{fig1}(c) we take the local adiabatic path by requiring $dB/dt\propto \Delta^2$ for $N=2$ ions \cite{monroe2021Programmable}. This allows us to dive into the adiabatic regime as much as possible within the available coherence time. As we can see, when $T\to 0$ the defect density is about 0.5, meaning that there is no correlation between the two spins in the initial state $|\uparrow_y\uparrow_y\rangle$. When the quench time $T$ increases, the defect density decreases, and finally if the adiabatic condition is satisfied in the whole process, we obtain a $\rho\propto T^{-2}$ scaling in the limit $T\to\infty$ (see inset). (In the experiment, the measured $\rho$ is lower-bounded by the detection infidelity of about 2-3\%, so we do not go to longer evolution time.) The KZM works in the middle of these two extreme cases where the system stays adiabatic except for a small region near the critical point with a small energy gap. Because the energy gap decays as we increase the ion number, our task becomes to find a shared scaling law for sufficiently large ion numbers, which does not require us to reach the fully adiabatic regime. Therefore in the following experiments we take the exponential quench path for convenience to save us from computing the energy gap for the multi-ion Hamiltonian along the path and from programming this numerical function into the experimental sequence.

\begin{figure}[!tbp]
	\centering
	\includegraphics[width=0.9\linewidth]{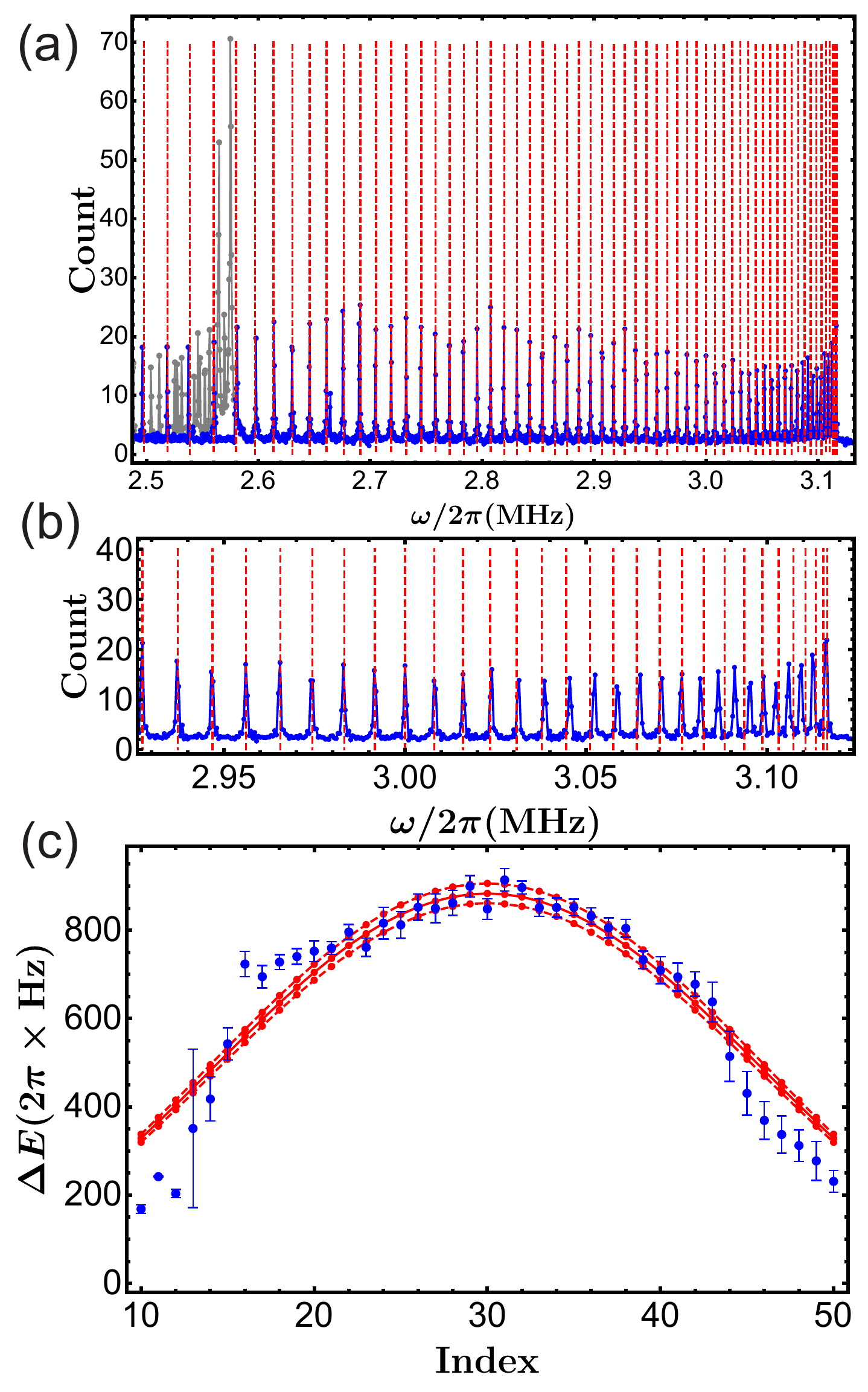}
	\caption {Calibration of the long-range Ising coupling.
(a) We scan the blue phonon sidebands for $N=61$ ions (blue dots) and use them to fit the inter-ion distances. (The grey dots are identified as peaks for the transverse modes in the orthogonal direction and are not relevant to this experiment.) The red dashed lines are the fitting results under a uniform transverse trapping frequency $\omega_x=2\pi\times 3.1166\,$MHz.
(b) A zoom-in of the fitted phonon band at the high-frequency end. The small discrepancy can be explained by the inhomogeneity in the transverse trapping potential over the $207\,\mu$m ion chain.
(c) We initialize the spin chain in $|\uparrow_x\cdots\uparrow_x\rangle$ ($|\uparrow_x\rangle\equiv (|\uparrow_z\rangle+|\downarrow_z\rangle)/\sqrt{2}$) and drive it with a weak oscillating transverse field term. The resonant frequencies for individual ions are shown as the blue dots with error bars representing one standard deviation, and are compared with the theoretically computed values (red dots). The upper and lower red dashed curves represent $\pm 2\pi\times 500\,$Hz shift in laser detuning. Raman $\pi/2$ pulses are used to initialize and to measure the spin states in the $\sigma_x$ basis. Due to the nonuniformity of the laser intensity over the long ion chain, such pulses are inaccurate for the edge ions, thus the corresponding results are not shown.
     \label{fig2}}
\end{figure}

\section{Critical behavior of FM long-range Ising model}
The universal behavior near the critical point is expected to be insensitive to local fluctuation of the parameters, but it can still depend on the decay rate of the long-range interaction \cite{PhysRevLett.109.267203,PhysRevB.94.075156,jaschke2017critical,puebla2019quantum}. Therefore, when studying the scaling with respect to the ion number, we need to ensure that the long-range Ising coupling $J_{ij}\approx J_0/|i-j|^\alpha$ maintains roughly the same $\alpha$. This requires us to first calibrate the coupling coefficients of the simulated Ising model.
In general, $O(N^2)$ cost is needed to calibrate all the coupling coefficients for $N$ ions, which is time-consuming and also requires individual addressing of the ions. Since the critical behavior is robust to small local errors, here we choose to calibrate the Rabi frequency and the collective phonon modes, compute the coupling coefficients theoretically as described below Eq.~(\ref{eq1}), and then take $O(N)$ cost to verify partial information about the coefficients.

As shown in Fig.~\ref{fig2}(a) and (b), we scan the phonon sidebands for $N=61$ ions to extract the frequency $\omega_k$ of each mode, and use them to fit the ion spacings. The overall spectrum fits well with the theoretical model apart from small deviation in the high-frequency end. For one thing, this is because we have about kHz frequency resolution for the measured $\omega_k$; for another, in the fitting model we assume a uniform transverse trap frequency $\omega_x=2\pi\times 3.1166\,$MHz over the chain given by the highest mode frequency (the center-of-mass mode), while in reality we find small variation of about $15\,$kHz in $\omega_x$ over the $207\,\mu$m ion chain. We further calibrate the Rabi frequency of the laser on each ion and then compute the theoretical Ising coupling coefficients $J_{ij}$ (see Appendix~\ref{app:parameters}).

To verify these results, we apply the coherent imaging spectroscopy method \cite{senko2014coherent}. We initialize the spins in $|\uparrow_x\cdots\uparrow_x\rangle$ and apply a Hamiltonian similar to Eq.~(\ref{eq1}) but with the transverse field term $B\ll |J_0|$ and oscillating at the frequency $\omega$. By scanning $\omega$, we look for the resonant frequency $\Delta E_i$ to flip the $i$-th spin, which gives us $\Delta E_i=2\sum_{j\ne i} J_{ij}$. As shown in Fig.~\ref{fig2}(c), the measured resonant frequencies agree well with the theoretical predictions for the central ions. For the ions near the edges, there are larger state-preparation-and-measurement (SPAM) errors due to the inhomogeneity in the Raman laser beams when converting between $\sigma_x$ and $\sigma_z$ bases, which may be improved by using composite pulses as we discuss later. Nevertheless, this is just the incapability to verify the coupling for the edge ions, and does not mean the inaccurate theoretical results. After such calibration and partial verification of the coupling $J_{ij}$, we can now fit $J_0$ and $\alpha$ for the long-range Ising interaction (see Appendix~\ref{app:theoretical} for more discussion about this fitting), and we can adjust the laser detuning $\delta$ to keep a constant $\alpha$ when we increase the ion number. This will allow us to examine how the model approaches the thermodynamics limit.

\begin{figure}[!tbp]
	\centering
	\includegraphics[width=1\linewidth]{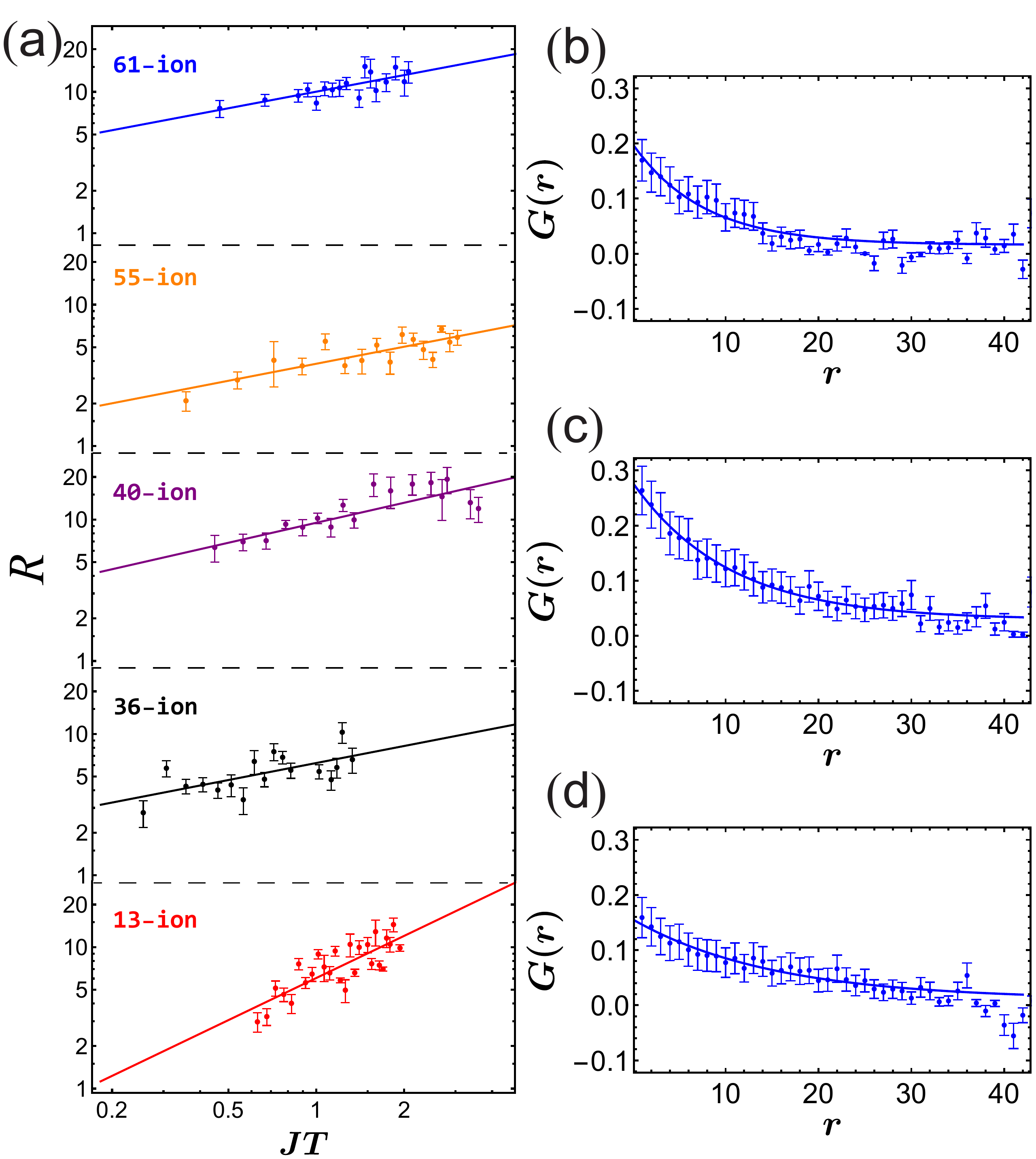}
	\caption {Quantum Kibble-Zurek mechanism and scaling. We initialize a polarized state $|\uparrow_y\cdots\uparrow_y\rangle$ along the transverse field and quench the field to zero following an exponential path $B(t)=B_0 [\exp(-t/\tau)-\exp(-T/\tau)]$ where $\tau=T/5$. To make the polarized state close to the initial ground state, we set a strong initial field $B_0=42.5 |J_0|$. For each total evolution time $T$, we measure the two-spin correlation $\langle \sigma_x^i \sigma_x^j\rangle - \langle\sigma_x^i\rangle\langle\sigma_x^j\rangle$ and further extract the correlation length $R$ on the ion chain through an exponential fitting. (See the right panels for $T=0.875\,$ms, $1.75\,$ms and $2.75\,$ms as some typical results at $N=61$. The reason for the different magnitudes of the correlation is discussed in Appendix~\ref{app:composite}.)
We summarize the results for various ion numbers $N$ vs. the dimensionless quench time $|J_0|T$ in the left panel as semi-log plots. For $N=\{13,36,40,55,61\}$, the slopes are
$\{0.99\pm 0.07,0.40\pm 0.08,0.47\pm 0.08,0.40\pm 0.06,0.39\pm 0.09\}$, respectively. Except for the smallest ion number $N=13$ where the finite size effect can be significant, we get consistent slopes as an estimation of the critical exponent $\mu\approx 0.42$.
    \label{fig3}}
\end{figure}

We present this scaling analysis for the KZM with various ion numbers under roughly the same $\alpha\approx 1$ in Fig.~\ref{fig3}. For each ion number $N$, we initialize the spin state in $|\uparrow_y\cdots\uparrow_y\rangle$ and ramp down the transverse field following an exponential path $B(t)=B_0 [\exp(-t/\tau)-\exp(-T/\tau)]$ for a total evolution time of $T$. Here $\tau$ is a parameter for the path and we fix $\tau=T/5$ for all the experiments. To suppress the artificial oscillation in Fig.~\ref{fig1}(c) due to the mismatch between the initial ground state and the fully polarized state we prepare, we choose a large initial field $B_0=42.5|J_0|$.
We measure the spatial spin correlation in the final state
\begin{equation}
G(r)\equiv \frac{1}{N_r} \sum_i (\langle\sigma_x^i \sigma_x^{i+r}\rangle - \langle\sigma_x^i\rangle\langle \sigma_x^{i+r}\rangle),
\end{equation}
and directly extract the correlation length $R$ from an exponential fitting $G(r)=A e^{-r/R} + B$ (see the right panels for some typical fitting results for $N=61$ ions). Here $N_r$ is the number of ion pairs with the distance $r$ and we discard the ions on the edges when computing the correlation length due to the larger SPAM errors described above. This correlation length should scale as $T^{\mu}$, from which we can obtain the critical exponent. The complete fitting results are summarized in the left panel for $N=13$, $36$, $40$, $55$ and $61$, and we get a consistent critical exponent $\mu\approx 0.42$ for all the ion numbers $N\ge 36$ which is also in agreement with the previous numerical result $\mu\approx 0.45$ \cite{jaschke2017critical}. The mismatch for the small ion number $N=13$ can be explained by the finite size effect. More discussion about the finite-size scaling and about the comparison to the theoretical results can be found in Appendix~\ref{app:theoretical}.

\begin{figure}[!tbp]
	\centering
	\includegraphics[width=1\linewidth]{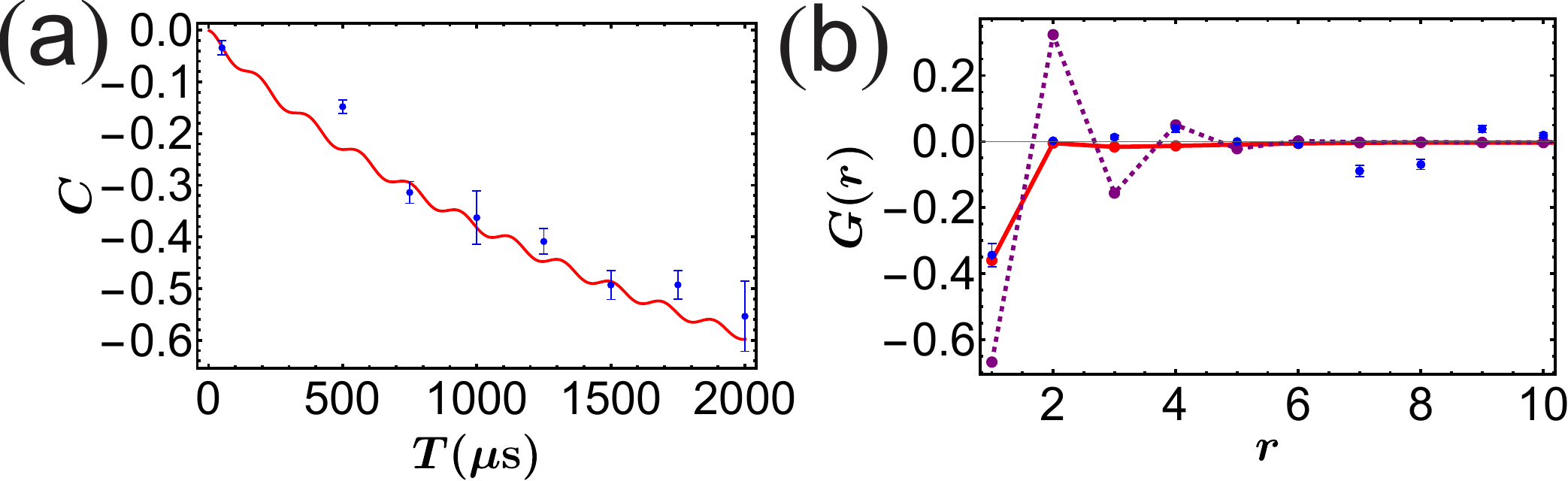}
	\caption {Similar experiments for anti-ferromagnetic Ising model. (a) For two ions, considerable change in the two-spin correlation from 0 to about -0.6 (which corresponds to a change in the defect density from 0.5 to 0.2) can still be observed as we increase the total quench time up to $T=5\tau=2\,$ms under the coupling $J_0=2\pi\times 265\,$Hz.
(b) However, for $N=13$ ions, the spatial correlation under $J_0 = 2\pi\times 67\,$Hz $\alpha=1.05$ and $T=5\tau=4\,$ms is much weaker (blue dots for experimental data and red solid line for theoretical prediction). This is due to the much smaller energy gap for the anti-ferromagnetic case. To observe significant spatial correlations, $20\,$ms total evolution time will be needed (purple dashed line).
    \label{fig4}}
\end{figure}

\section{AFM long-range Ising model}
Similar methods can also be applied to the AFM Hamiltonian. Specifically, we initialize the spin state in $|\downarrow_y \cdots \downarrow_y\rangle$, which is close to the ground state rather than the highest excited state of the initial Hamiltonian under a strong transverse field. Then we follow the same exponential path $B(t)=B_0 [\exp(-t/\tau)-\exp(-T/\tau)]$ to turn down the transverse field and go across the phase transition point for a total evolution time of $T$, and again we measure the spin-spin correlation $\langle\sigma_x^i\sigma_x^j\rangle$ of the final state.
However, in this case due to the frustration in the interaction, the energy gap becomes much smaller and thus we expect a much weaker signal under the same quench time. As shown in Fig.~\ref{fig4}(a), for two ions we still observe significant change in the spin-spin correlation as we tune the quench time from zero to $2\,$ms. (Indeed, theoretically one can see that in this case the correlation is exactly opposite to that in the FM case.) However, if we go to larger ion numbers such as $N=13$ in Fig.~\ref{fig4}(b), the correlation quickly decays from $r=1$ to $r=2$, which prevents us from fitting a correlation length. Theoretically we can also simulate the 13-spin dynamics numerically and we get consistent results. To observe strong correlations, we may need $20\,$ms total evolution time, for which the other decoherence effects cannot be neglected.

\section{Discussion}
To sum up, in this work we experimentally realize the long-range transverse-field Ising model and examine its critical behavior using the KZM. We vary the system sizes while maintaining the same long-range interaction to obtain a critical exponent after suppressing the finite size effect. Our experiment goes beyond the qualitative understanding of the simulated physical system and makes a quantitative application of the ion trap quantum simulator.

For this long-range interacting model to be well-defined in the thermodynamic limit, usually Kac normalization needs to be considered. For example, if we want to compare the phase transition point for different system size $N$, then we need to rescale the coupling strength $J_{ij}$ by the Kac normalization $\frac{1}{N-1} \sum_{i\ne j}\frac{1}{|i-j|^\alpha}$ (similar expressions can be found in, e.g., Ref.~\cite{PhysRevB.96.104436}). However, in this experiment we are measuring the critical exponent through the quantum Kibble-Zurek mechanism, so that the detailed phase transition points for different $N$ are not important so long as we always start the quench deep in one phase and end it deep in the other phase. This is ensured by a large $B_0=42.5|J_0|$ in Fig.~\ref{fig3}, which is sufficient even for the largest system size $N=61$ we use.

Currently our spin system has a coherence time of several milliseconds under the laser driving, which is fitted from the two-ion Ising model dynamics and mainly comes from the slow drift in the laser intensity among different experimental trials.
This is still below the reported coherence time of tens of milliseconds for two-qubit gates \cite{PhysRevLett.125.150505} and may be improved by better stabilization of the system. However, note that in this experiment such a slow drift can be regarded as a relative change between $J_{ij}$'s and $B_i$'s and a rescaling in the quench time. Different from the fast oscillation in the Ising dynamics, here the KZM will not be significantly affected so long as the initial condition $B_0 \gg |J_0|$ is still satisfied.
Another possible error source is the motional decoherence, again on the order of several milliseconds, which is measured from the Ramsey experiment on the blue motional sideband. However, in this experiment, to generate the long-range transverse-field Ising model Hamiltonian in Eq.~(\ref{eq1}) we keep the phonon modes to be only virtually excited, so that its effect is also suppressed. Specifically, we use a Raman Rabi frequency of about $2\pi\times 100\,$kHz and a detuning to the nearest phonon sideband of about $2\pi\times 20\,$kHz, such that the phonon excitation per ion is estimated to be on the order of 1\% given a Lamb-Dicke parameter $\eta_k\sim 0.1$. This should suppress the effect of motional decoherence on the spin system by about 100 times.
Other sources of decoherence such as the phase fluctuation between the Raman laser beams (measured from the Ramsey experiment) and the spontaneous emission (computed from the off-resonant excitation to the $\mathrm{P}_{1/2}$ levels) also have a coherence time above $100\,$ms, much longer than our largest quench time below $5\,$ms.

In this work we set the detuning for large coupling $|J_0|$ and get $\alpha\approx 1$. With an enhanced evolution time in the future, it will be possible to choose other $\alpha$ at the cost of smaller $|J_0|$, thus allowing us to explore the dependence of the critical exponent on the long-range interaction $\alpha$. Also, for a long ion chain the inhomogeneity of the laser becomes important and can cause considerable SPAM errors for the edge ions. In this experiment this may not be a significant problem because we only use the central ions with large spin-spin correlations to fit the correlation length, but we can also use composite pulses \cite{PhysRevA.70.052318} to implement such $\pi/2$ rotations for better performance as described in Appendix~\ref{app:composite}.

\begin{acknowledgments}
This work was supported by Innovation Program for Quantum Science and Technology (2021ZD0301601), the Tsinghua University Initiative Scientific Research Program and the Ministry of Education of China through its fund to the IIIS. Y.-K. W. acknowledges support from the start-up fund from Tsinghua University.
\end{acknowledgments}

\appendix
\section{Experimental setup and calibration of parameters}
\label{app:parameters}
We use a segmented blade trap to confine $^{171}\mathrm{Yb}^{+}$ ions. We perform Doppler cooling followed by sideband cooling before each experiment trial to cool the phonon number in each mode to be below 0.1. The daily operations of the ions and the methods to calibrate the experimental parameters are the same as those in our previous work \cite{RabiHubbard}.

\begin{figure*}[!tbp]
	\centering
	\includegraphics[width=0.6\textwidth]{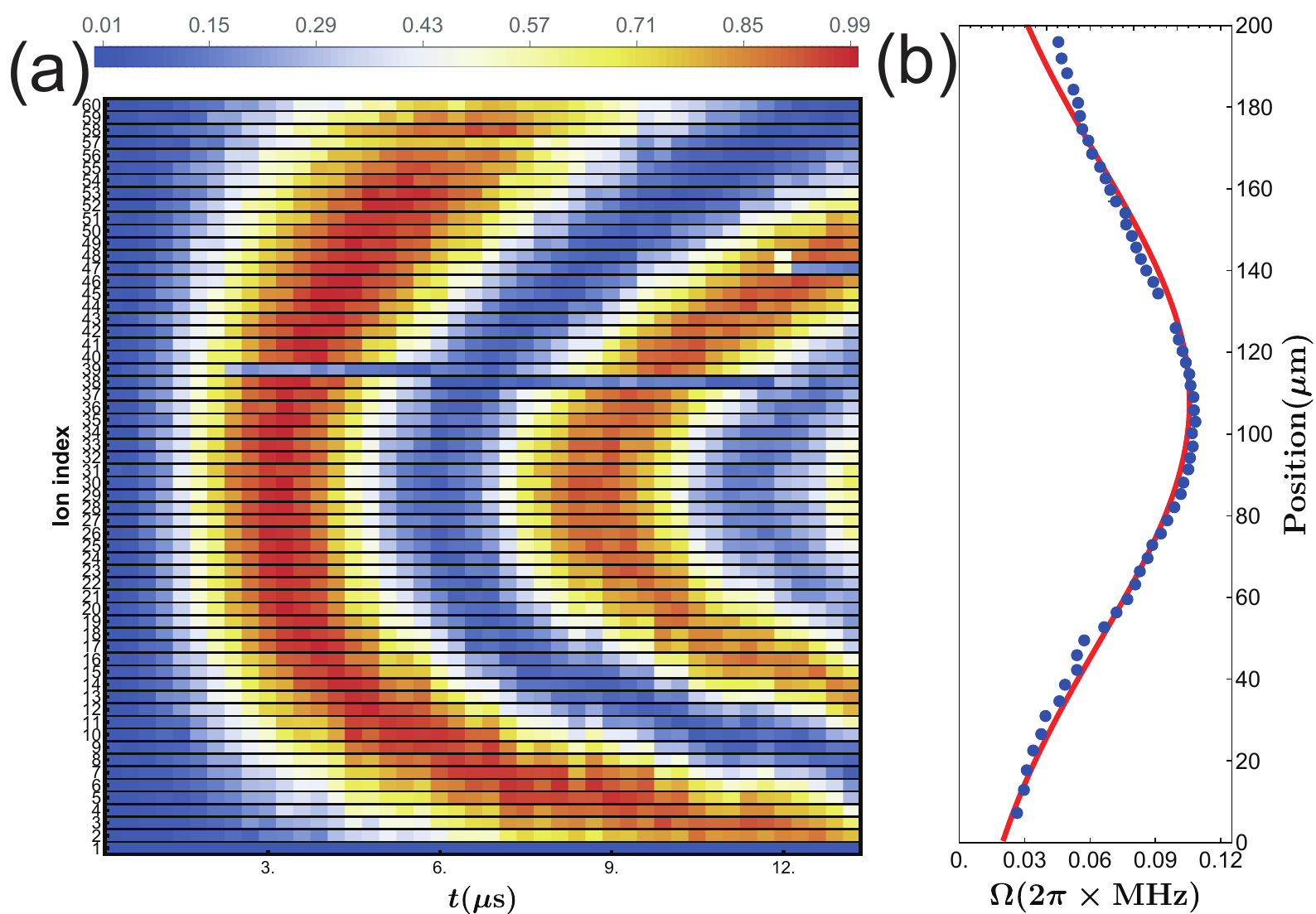}
	\caption {Calibration of Rabi frequency. Here we set a total laser power of $800\,$mW for the two counter-propagating beams. The Rabi frequency under other laser intensities can be scaled accordingly. (a) Rabi oscillation of $N=60$ ions. The missing data are due to two dark ions in the calibration process (likely being on the $F_{7/2}$ levels or becoming $\mathrm{Yb H^+}$). (b) The Rabi frequency is fitted by a Gaussian distribution with a full width at half maximum of $\sim 144\,\mathrm{\mu}$m.
    \label{figS1}}
\end{figure*}

We calibrate the Rabi frequency of the global Raman laser on each ion as shown in Fig.~\ref{figS1}. The global Raman laser makes an equal angle of $45^{\circ}$ to the transverse $x$ and $y$ directions. We set the splitting between these two directions to be $\sim 300\,$kHz for 2 and 13 ions, and $\sim 550\,$kHz for 36-61 ions, and choose the Raman laser detuning to be slightly above the higher one $\omega_x$ (typically $10$-$20\,$kHz), such that the off-resonant coupling to the undesired $y$ modes can be safely neglected.

\begin{figure*}[!tbp]
	\centering
	\includegraphics[width=\textwidth]{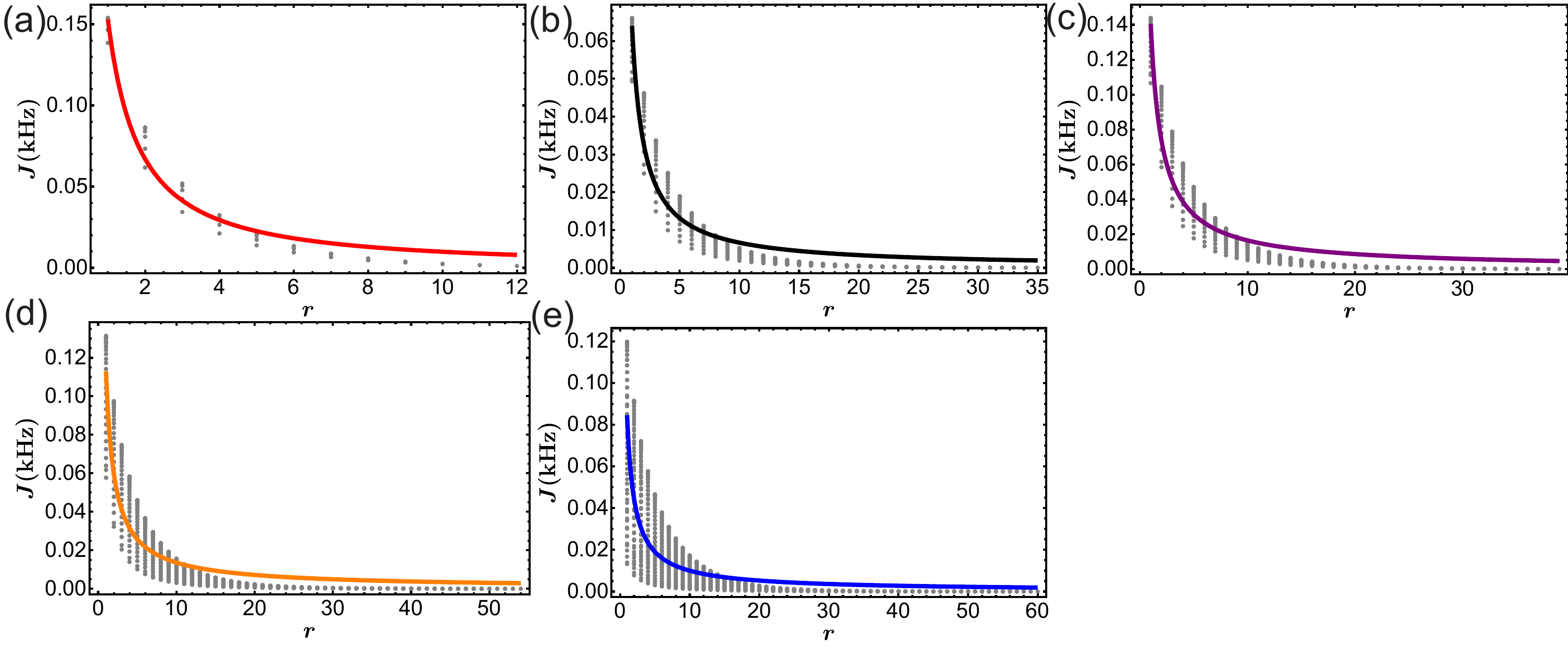}
	\caption {Fit long-range Ising coupling. (a)-(e) Fitting results for $N=13,36,40,55,61$ ions, respectively. Grey dots are the numerical results from the calibrated parameters. Solid lines are the $J_0/r^{\alpha}$ fitting results.
    \label{figS2}}
\end{figure*}

After calibrating the ion spacings, we compute the collective normal modes and the Ising coupling $J_{ij}=\Omega_i \Omega_j \sum_k \eta_k^2 b_{ik}b_{jk}\omega_k/(\delta^2-\omega_k^2)$. Then we further fit them by the power law $J_{ij}\approx J_0/|i-j|^\alpha$. The fitting results for the coupling used in Fig.~\ref{fig3} are shown in Fig.~\ref{figS2}. Note that the difference in $J_0$ can be rescaled by the quench time $T$. What matters is to keep $\alpha$ roughly the same for different system sizes. For $N=\{13,36,40,55,61\}$ ions, we get $J_0=2\pi\times\{153\pm 2,64\pm 1,141\pm 2,113\pm 1,83\pm 1\}\,$Hz and $\alpha=\{1.19\pm 0.03,0.98\pm 0.01,0.93\pm 0.01,0.92\pm 0.01,0.87\pm 0.02\}$, respectively. Observe that the fitting model is not perfect, but works reasonably well for the nearby ions.

When measuring the correlation length in the KZM, we discard $\{1,4,4,8,8\}$ ions on each end in the $N=\{13,36,40,55,61\}$ case to suppress the boundary effect.

\section{Composite pulses}
\label{app:composite}
\begin{figure}[!tbp]
	\centering
	\includegraphics[width=\linewidth]{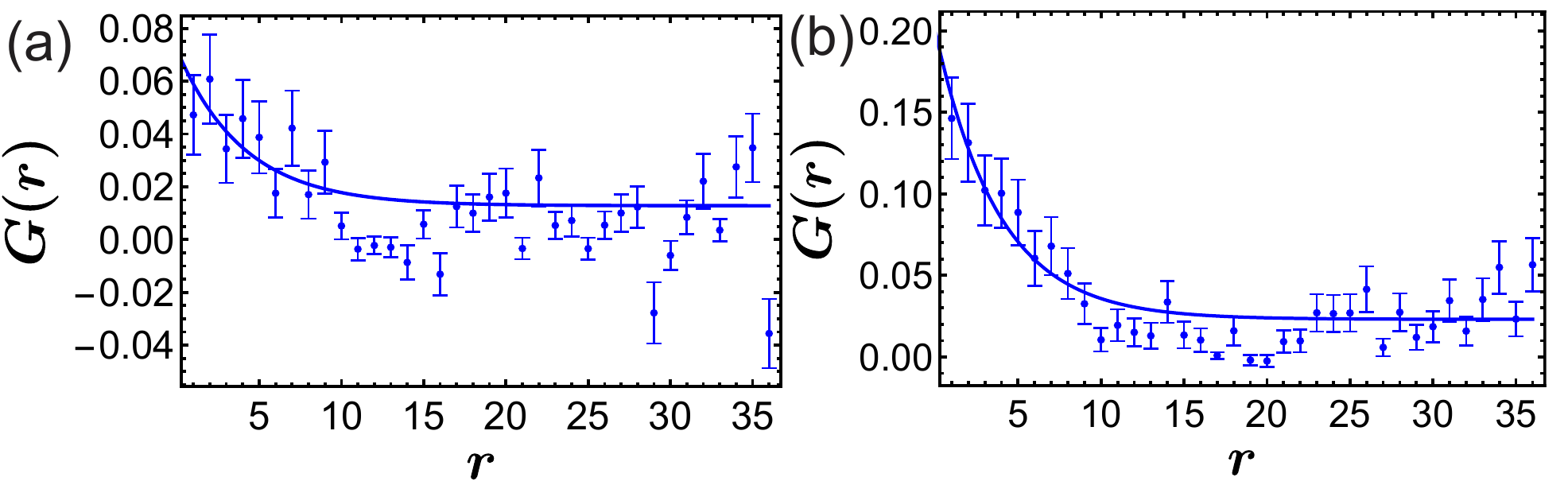}
	\caption {State preparation and measurement through composite pulses. (a) Measured spin-spin correlation using single pulses for $N=55$ and $T=4$ ms, from which we fit $R=4.0\pm 1.4$. (b) The same experiment using composite pulses, which gives $R=3.8\pm 0.8$. The composite pulse method gives smoother results at large distance where the inhomogeneity of the laser intensity becomes important.
    \label{figS3}}
\end{figure}

For long ion chains, the nonuniform Rabi frequency can lead to SPAM errors in the $\sigma_x$ or $\sigma_y$ bases for the ions on the edge. It can be improved by using composite pulses \cite{PhysRevA.70.052318} such as the BB1 pulse. Nevertheless, we find that this has little influence to the measured correlation length in this experiment, which is dominated by the central ions with strong correlation. In Fig.~\ref{figS3} we compare the experimental results for a typical quench time $T=4\,$ms for $N=55$ ions. The fitted correlation lengths are very similar with and without the use of composite pulses.

Meanwhile, we observe that the absolute values of the measured correlations differ in the two plots. This can be caused by an inaccurate setting of the bichromatic laser frequency ($\omega_b$ and $\omega_r$ in Fig.~\ref{fig1}, because the Ising model Hamiltonian in Eq.~(1) is achieved in an interaction picture rotating at the frequency $(\omega_b+\omega_r)/2$ \cite{RabiHubbard}. Although the errors in these frequencies can be controlled to be below $100\,$Hz, they can still accumulate into considerable misalignment in the $\sigma_x$ measurement after an evolution time of several milliseconds. As we describe in our previous work \cite{RabiHubbard}, this can be solved by scanning the measurement basis on the $\sigma_x$-$\sigma_y$ plane to extract the oscillation amplitude as the spin-spin correlation. However, since it is a global effect for all the spins, it will not influence our measurement of the spatial correlation length.

\section{Comparison with theoretical results}
\label{app:theoretical}
\begin{figure}[!tbp]
  \centering
  \includegraphics[width=0.9\linewidth]{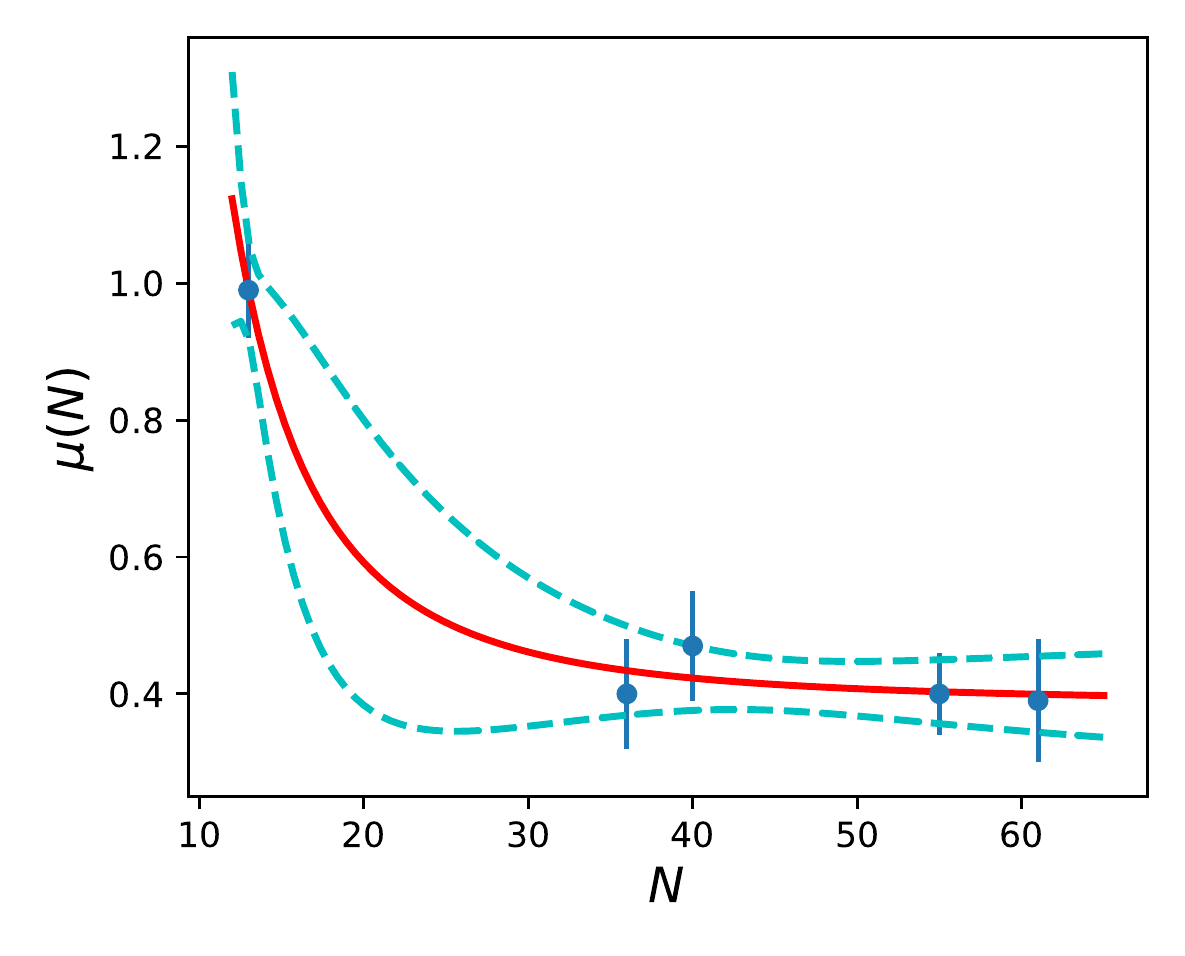}\\
  \caption{Fitting $\mu(N)$ with $\mu_{\infty}+a N^{-b}$. The best fit is given by $\mu_\infty=0.39\pm 0.11$, $a=(0.4\pm 2.9)\times 10^{3}$, $b=2.5\pm 3.1$ as the red curve. The upper and lower dashed curves represent one standard deviation.}\label{fig:finite_size_scaling}
\end{figure}

In Fig.~\ref{fig3}, we measure the slopes for different system sizes $N=\{13,36,40,55,61\}$ as
$\mu=\{0.99\pm 0.07,0.40\pm 0.08,0.47\pm 0.08,0.40\pm 0.06,0.39\pm 0.09\}$. Then in principle we should perform finite size scaling to extrapolate the critical exponent $\mu_{\infty}$ in the thermodynamic limit $N\to\infty$ by fitting $\mu(N)=\mu_{\infty}+a N^{-b}$ where $a$ and $b$ are two fitting parameters \cite{jaschke2017critical}. As shown in Fig.~\ref{fig:finite_size_scaling}, this gives us $\mu_{\infty}=0.39\pm0.11$ with a relatively large error bar. Another way to understand this result is that, for $N\ge 36$, the fitted slopes already agree with each other within the error bars, so that it becomes difficult to further extrapolate to $N\to\infty$ and the final result stays close to their average $\bar{\mu}=0.42$ with a relatively large error bar. This is the value we use in the main text.

The above measured critical exponent agrees well with the numerical result $\mu\approx 0.45$ in Ref.~\cite{jaschke2017critical}. However, we also note that there are analytical results suggesting that the critical exponent $\mu\equiv\nu/(1+z\nu)$ should diverge as $1/(\alpha-1)$ near $\alpha\approx 1$ \cite{PhysRevB.64.184106}. This difference could arise from the different methods used to extract the critical exponent, whether from the ground state \cite{PhysRevB.64.184106} or from the quench dynamics \cite{jaschke2017critical}. Another possibility is that, since the long-range interaction is of interest, one may need larger system sizes than those for the nearest-neighbor cases to reliably extrapolate the critical exponent $\mu_\infty$.

Finally, note that in Fig.~\ref{figS2} the fitting of $J_{ij}$ to a power-law decay is not perfect. This is common for ion trap experiments as the $1/|i-j|^\alpha$ scaling is only an approximation, and in this experiment it is intensified by the nonuniform laser intensity over the long ion chain. In Appendix~\ref{app:parameters} we perform the least square regression for all the $J_{ij}$'s, but since there are more ion pairs with small distances, the fitting is not as good for the pairs with larger distances (see the tails of Fig.~\ref{figS2}). Overall, as the correlation length diverges around the phase transition point, we expect the critical exponent to be insensitive to the small fluctuation around the power-law scaling, but whether this globally fitted $\alpha$ or a larger one with higher weights on the distant pairs can better represent this long-range Ising model, still remains an open question for future research.

%

\end{document}